# Prediction of a Novel 2D Porous Boron Nitride Material with Excellent Electronic, Optical and Catalytic Properties

*Vikram Mahamiya[a], Alok Shukla[a*], and Brahmananda Chakraborty[b,c]*

[a]Indian Institute of Technology Bombay, Mumbai 400076, India

[b]High pressure and Synchrotron Radiation Physics Division, Bhabha Atomic Research Centre, Bombay, Mumbai, India-40085

[c]Homi Bhabha National Institute, Mumbai, India-400094

email: mahamiyavikram@gmail.com ; shukla@iitb.ac.in ; brahma@barc.gov.in

**ABSTRACT**

Holey graphyne (HGY) is a recently synthesized two-dimensional semiconducting allotrope of carbon composed of a regular pattern of six and eight-vertex carbon rings. In this study, based on first-principles density functional theory and molecular dynamics simulations, we predict a similar stable porous boron nitride holey graphyne-like structure that we call BN-holey-graphyne (BN-HGY). The dynamical and thermal stability of the structure at room temperature is confirmed by performing calculations of the phonon dispersion relations, and also *ab-initio* molecular dynamics simulations. BN-HGY structure has a wide direct bandgap of 5.18 eV, which can be controllably tuned by substituting carbon, aluminum, silicon, and phosphorus atom in place of sp and $sp^2$ hybridized boron and nitrogen atoms of BN-HGY. We have also calculated the optical properties of the HGY and BN-HGY structures for the first time and found that the optical absorption spectra of these structures span full visible and a wide range of ultraviolet regions. We found that the Gibbs free energy of the BN-HGY structure for the hydrogen adsorption process is very close to zero (-0.04 eV) and, therefore, the BN-HGY structure can be utilized as a potential catalyst for HER. Therefore, we propose that the boron nitride analog of holey graphyne can be synthesized, and it has a wide range of applications in nanoelectronics, optoelectronics, spintronics, ultraviolet laser, and solar cell devices.

## 1. INTRODUCTION

Carbon nanomaterials have received a lot of scientific interest due to the ability of carbon to get hybridized in $sp^3$, $sp^2$, and sp form and making various stable allotropes, with unique electronic properties[1–4]. The experimentally synthesized carbon allotropes of various dimensions, including fullerenes[5], carbon nanotubes[6], graphene[7] etc. have been explored widely. Pristine graphene has zero energy bandgap that limits its applications in the operation of electronic and optoelectronic devices. Therefore, a lot of scientific research has been carried out for the bandgap opening of graphene and synthesizing other 2D materials with suitable bandgaps. There has been extensive study to open the bandgap of graphene by using defects, doping, external electric field, strain, changing the stacking nature, etc.[8–15]. In addition to that, there are some other semiconducting nanostructures, including transition metal dichalcogenides ($MX_2$, where M = Mo, W; X = S, Se)[16–20], Mxenes[21–24], hexagonal boron nitrides (h-BN)[25,26], etc., that have been thoroughly explored due to their applications in electronics, photonics, and optoelectronics. Graphyne family structures which were formed by benzene rings and acetylene linkage have gained enormous scientific interest due to their large holes, which can be utilized for energy storage purposes[27–29]. By increasing the length of the acetylene linkage in graphyne structures, the electronic bandgap can be tuned, and it was found that the graphyne family has exceptional electronic[30], optical[31], and mechanical properties[32]. Furthermore, the carrier mobility in graphyne is even more than graphene[33].

After the experimental synthesis of graphyne structures in 2017[34,35], their electronic, optoelectronic, and spintronic properties have been widely explored. The graphyne family structures are utilized as anodes for battery applications[36–38], transistors[39,40], gas sensing[41–44], hydrogen evolution reaction and storage[45–50], etc. Wang *et al.*[51] have synthesized boron-graphdiyne structure and explored its application for Na ion storage. Ding *et al.*[52] have recently synthesized N-graphyne structure by one-step ball mining of $CaC_2$ and pyrazine. They have reported that N-graphyne has potential applications in electrocatalysis and supercapacitor. The graphitic carbon $g-C_3N_4$ is another two dimension structure which has been thoroughly explored after its experimental realization[53,54]. The nitrogenated holey graphene ($C_2N$) structure has also been synthesized and explored for field-effect transistor applications by Mahmood *et al.*[55]. The experimental synthesis of these carbon-boron and carbon-nitrogen structures has opened the door towards post-silicon nanoelectronics. The controlled introduction of boron and nitrogen atoms into carbon nanostructures can tune their electronic and chemical properties. The BN is isoelectronic to CC, having a similar bond length but very different electronic properties. The BN analogs of carbon nanostructures have a wide bandgap and have potential applications in nanoelectronics device fabrication[56–59]. The two-dimensional BN sheets, BN analog of graphyne BN-yne and BN nanoribbons have been synthesized and theoretically studied[60–62]. The BN analog of two-dimensional porous triphenylene graphdiyne (TpG) has also been theoretically explored by Muhammad *et al.*[63].

Very recently, a new two-dimensional porous carbon structure with a periodic pattern of six and eight-vertex rings, holey graphyne (HGY), has been successfully synthesized[64]. Liu *et al.*[64] have reported that HGY is a p-type semiconductor with high electron and hole mobility at room temperature. The HGY structure has uniform periodic holes, which can be utilized for energy storage applications. Hydrogen storage in Li and Sc decorated HGY structures have been explored by Guo et al.[65] and Mahamiya et al.[66], respectively.

In this paper, we have explored the possibility of synthesizing BN analog of two-dimensional HGY by theoretically investigating its stability and various properties suitable for different applications. We found that the BN analog of HGY structure is energetically and dynamically stable. We have also observed that the bandgap of the BN-HGY structure can be tuned, and that the optical absorption of HGY and BN-HGY span a wide range of visible and ultraviolet regions. In addition to that, the BN-HGY structure exhibit excellent catalytic activity towards hydrogen evolution reaction (HER). Hence, the proposed BN-HGY system has potential applications in nanoelectronics, optoelectronics, solar cells, and hydrogen production.

## 2. COMPUTATIONAL DETAILS

We have performed the density functional theory (DFT) and *ab-initio* molecular dynamics (AIMD) calculations using the Vienna *ab-initio* simulation package (VASP)[67–70]. The generalized gradient approximation (GGA) with Perdew-Burke-Ernzehof (PBE) exchange-correlation functional was used for the calculations[71]. We have taken a vacuum space of 20 Å along the z-direction to avoid the periodic interaction between two consecutive layers. A Monkhorst-pack K-point grid 5×5×1 was taken to sample the Brillouin zone, and convergence limits of 0.01 eV/ Å and $10^{-5}$ eV were considered for the force and total energy optimization, respectively. A Monhorst-pack grid of 7×7×1 kpoints was taken to calculate the electron localization function (ELF)[72] for the BN-HGY structure. The ELF can explain the qualitative distribution of electron density and the nature of chemical bonds present in this novel material. The values of ELF ranges between zero to one, corresponding to a region of very low electron density to high electron density for localized electrons, respectively. The phonon dispersion calculations were performed using density functional perturbation theory method. We have checked the stability of the structure at room temperature by performing the *ab-initio* molecular dynamics (AIMD) simulations. AIMD simulations solve electronic Schrodinger equations for the nuclei at each time step and compute the potential energy surfaces. Thus a lot of crucial information, such as charge transfer, bond dissociation and formation, change in electronic states[73,74], etc., can be obtained by performing AIMD simulations which are difficult to find out from the classical MD based on empirical force fields. However, AIMD simulations are very costly since each time step involves a single DFT energy and force calculations, so trajectories are limited to picoseconds of simulation time, and the general time step should be of the order of femtoseconds to ensure numerical stability.[75] The AIMD simulations are performed in two consecutive steps, initially, we put the relaxed structure of BN-HGY in a microcanonical ensemble (NVE) for 5 ps time and the temperature of the system is increased up to 300 K in a time step of 1 fs. Next, we kept this structure in a canonical ensemble (NVT) at 300 K by employing the Nose–Hoover thermostat[76] for another 5 ps time duration. We have also investigated the thermal stability of the BN-HGY structure at a high temperature of 1000 K by following the same procedure.

## 3. RESULTS AND DISCUSSION

### 3.1 Structural stability

The final optimized structures of the 2×2×1 supercell of HGY and BN-HGY are shown in **Fig. 1 (a & b).** The lattice constants for the hexagonal BN-HGY structure are a = b = 10.92 Å, which are close to the pristine HGY lattice constants value of 10.85 Å[65]. There are four different values of BN bond lengths corresponding to $sp^2$-$sp^2$, $sp^2$-sp, and sp-sp hybridized boron and nitrogen atoms, as shown in **Fig. 1 (b).** C1, B1, and N1 denote sp-hybridized carbon, boron, and nitrogen atoms, while C2, B2, and N2 denote sp$^2$-hybridized carbon, boron, and nitrogen atoms, as displayed in **Fig. 1 (a & b).**

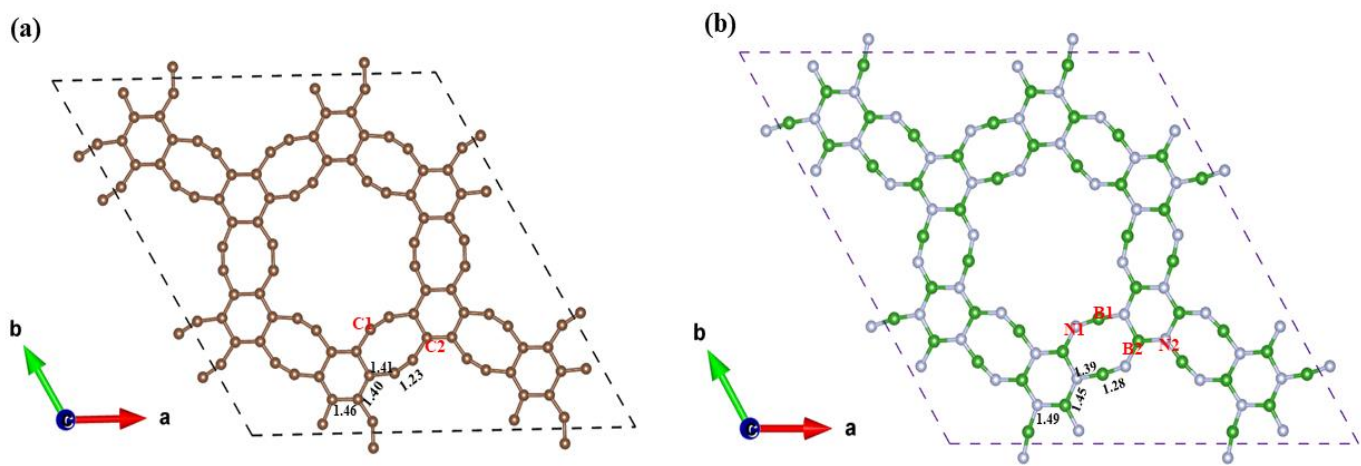

**Fig. 1 Optimized structure of (a) 2×2×1 supercell of HGY. (b) 2×2×1 supercell of BN-HGY. Here C1, B1 and N1 represent sp-hybridized carbon, boron and nitrogen atoms, while C2, B2 and N2 represent sp$^2$ hybridized carbon, boron and nitrogen atoms, respectively.**

The BN-HGY structure has two sp$^2$-sp$^2$ hybridized bonds with bond lengths 1.49 and 1.45 Å, and one each sp$^2$-sp and sp-sp hybridized bond with bond lengths 1.39 and 1.28 Å, respectively. Since BN and CC are isoelectronic and the boron and nitrogen atoms are located on sites similar to those of the carbon atoms in HGY, the change in the bond lengths of the BN analog of HGY compared to HGY is small. The change in the CC and BN bond length arises because the atomic radii and the electronegativities of boron and nitrogen atoms are different as compared to the carbon atoms. The angle between the hexagon of BN-HGY and acetylene linkage is 125.1°, which is close to the corresponding bond angle 126.03° in HGY, and larger than the bond angle 120° for graphyne. A detailed comparison of the structural parameters of HGY and BN-HGY is presented in **Table 1.**

**Table 1. The structural parameters of HGY and BN-HGY.**

| Structural parameter | Bond lengths (Å) | | | Bond angle (degree) |
|---|---|---|---|---|
| | $sp^2 - sp^2$ | $sp^2 - sp$ | $sp - sp$ | |
| **HGY** | 1.46, 1.40 | 1.41 | 1.23 | 126.03 |
| **BN-HGY** | 1.49, 1.45 | 1.39 | 1.28 | 125.10 |

We have investigated the stability of the BN-HGY structure by calculating its cohesive and per-atom energy. We have also calculated the cohesive and per-atom energy of some stable BN analogs including, the BN analog of graphene and BN analog of carbon nanotube, to compare with the BN-HGY structure and investigated the feasibility of its synthesis. The cohesive energy of the BN-HGY structure is found to be -6.52 eV which is very close to the cohesive energy of the HGY structure -6.76 eV calculated at the same level of theory. Shin *et al.*[77] have found that the cohesive energy of the most stable graphyne form γ-graphyne is -6.76 eV, very similar to HGY and BN-HGY structures. The cohesive energy of monolayer $MoS_2$ is -4.98 eV[78] lesser than the cohesive energy of BN-HGY. We have found that the cohesive energies of the BN analogs of graphene and carbon nanotube are found to be -7.12 eV and -6.96 eV, respectively. Further, we have also calculated that the per-atom energy of the BN-HGY structure -8.21 eV, which is also very close to the per-atom energy of the HGY structure, -8.46 eV. The per-atom total energies of the BN analogs of graphene and carbon nanotube are found to be -8.81 eV and -8.66 eV, respectively. The cohesive and per-atom energies of the BN analog of HGY are very close to those quantities of the pristine HGY and other stable BN analog structures, suggesting that the experimental realization of the BN-HGY structure is energetically feasible.

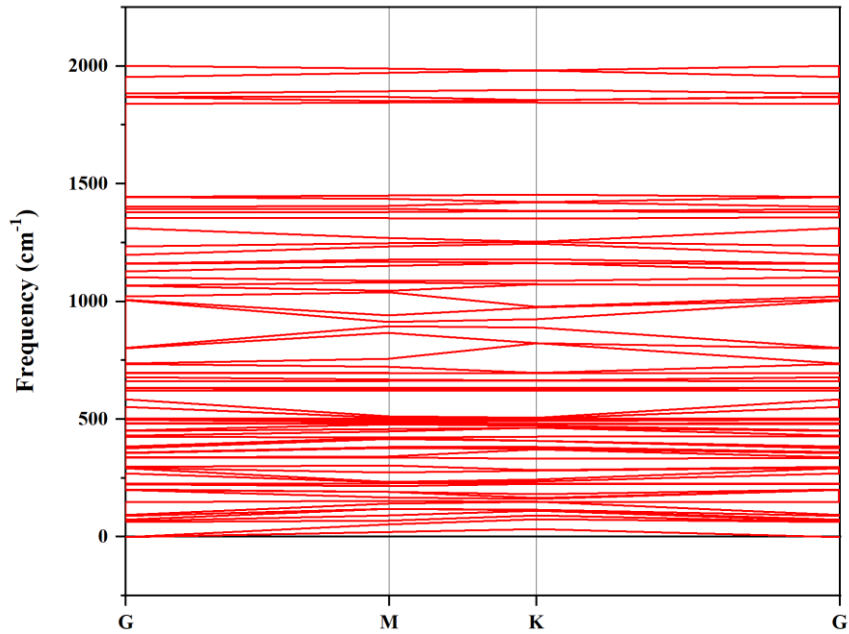

**Fig. 2 Calculated phonon dispersion relations of BN-HGY.**

The phonon dispersion curve of the BN-HGY presented in **Fig. 2,** contains only real frequencies in the full Brillouin zone, indicating that the BN-HGY structure is dynamically stable.

### 3.2 Thermal stability

To check the thermal stability of the structure at room temperature and also at high temperatures, we have performed *ab-initio* molecular dynamics simulations. The molecular dynamics snapshot of the BN-HGY after keeping it in a canonical ensemble for 5 picoseconds at room temperature is shown in **Fig. 3(a).** The thermal fluctuations around room temperature are small, as shown in **Fig. 3(b).** We found that the structure does not deform at room temperature and the changes in the B-N bond lengths are negligible. The bond length fluctuations of the sp-sp, sp-sp$^2$, and sp$^2$-sp$^2$ B-N bonds are small at room temperature, as shown in **Fig. 4(a)**. We have observed that the maximum fluctuations for sp-sp, sp-sp$^2$, and sp$^2$-sp$^2$ bonds are 3.9 %, 7.7 %, and 6.2 % of their mean values, respectively, at room temperature. Although the maximum fluctuations in B-N bonds increase up to ~ 20 % at a high temperature of 1000 K, as shown in **Fig. 4(b)**, the B-N bonds do not break, and the structure is intact.

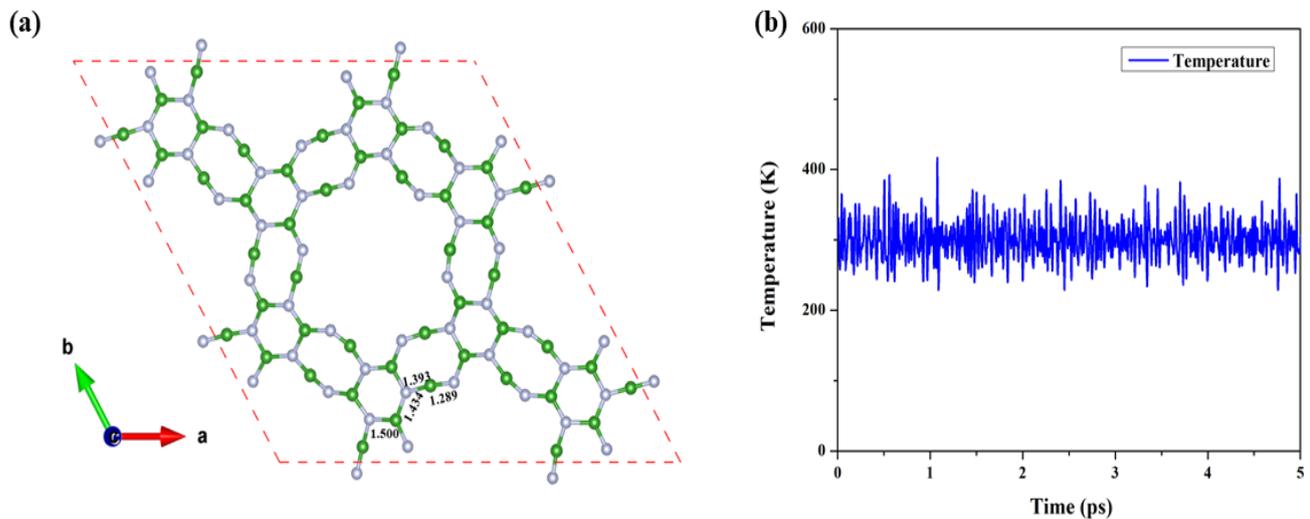

**Fig.3 (a) Molecular dynamics snapshot of BN-HGY. (b) Fluctuations of temperature around room temperature.**

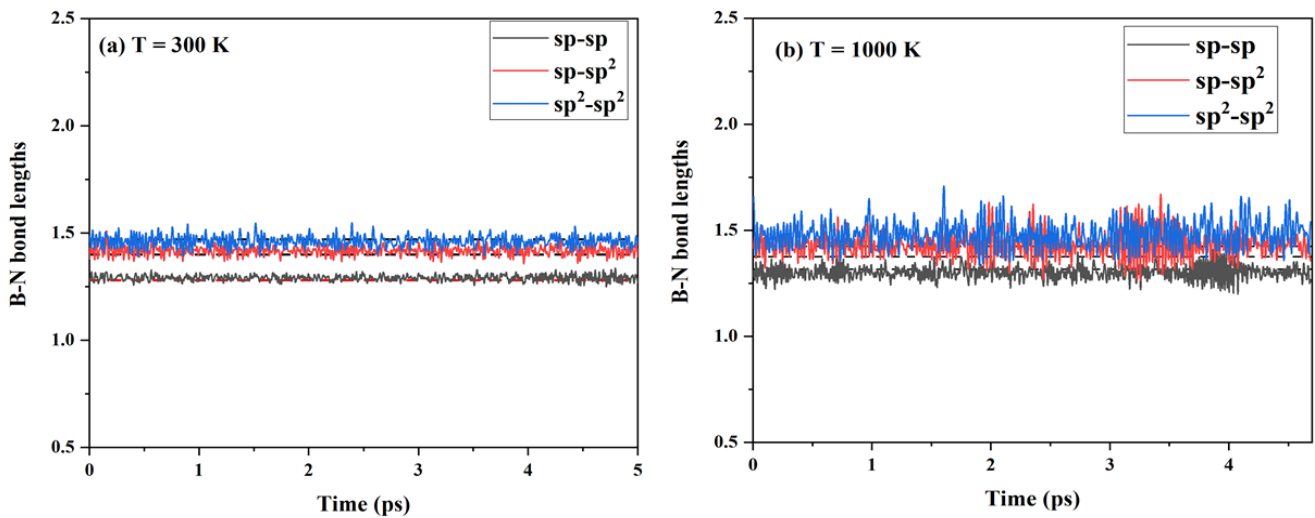

Fig. 4 Fluctuations in sp-sp, sp-sp$^2$, and sp$^2$-sp$^2$ hybridized B-N bond lengths at (a) 300 K, (b) 1000 K, after keeping the system in a canonical ensemble for 5 ps time duration.

The molecular dynamics snapshot and structural coordinates of BN-HGY, after putting the structure in a canonical ensemble for 5 picoseconds time duration at 1000 K, are provided in the supporting information file (See **Fig. S1**). The B-N bonding in the two-dimensional BN-HGY structure is very strong due to which BN-HGY material can withstand up to high temperatures. Previously, Muhammad *et al.*[63] have reported the BN-analog of two-dimensional triphenylene-graphdiyne is stable up to 1500 K, and the structure breaks at around 2000 K by performing AIMD simulations for 5 ps in the time step of 1 fs. The thermal fluctuations in the BN-HGY structure, after keeping the system in a canonical ensemble at 1000 K for 5 ps time are moderate, as presented in **Fig. S2** of the supporting information file.

### 3.3 Electronic properties

The total density of states of HGY and BN-HGY structures are shown in **Fig. 5.** The pristine HGY structure has a direct bandgap of 0.4 eV when calculated using PBE exchange-correlation functional, which is in excellent agreement with the literature.[79,80]

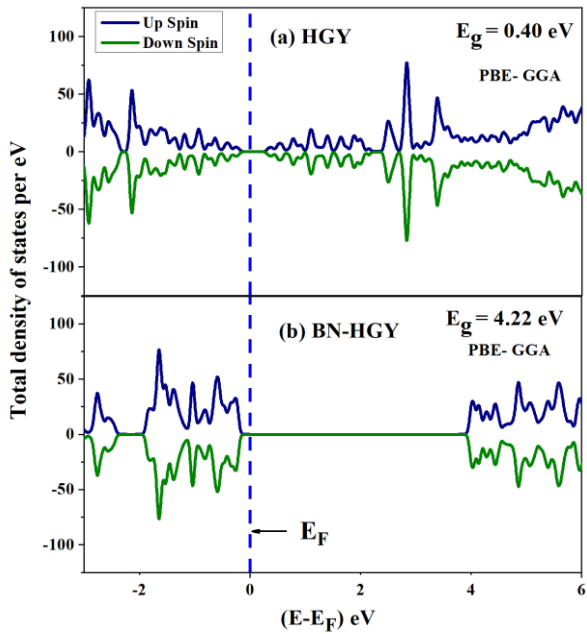

**Fig. 5 Total density of states of (a) Monolayer holey graphyne (b) Boron nitride analog of monolayer holey graphyne. $E_F$ represents the Fermi level.**

On the other hand, using the same functional, the band gap of BN-HGY also turns out to be direct, but with the value 4.22 eV, which is comparatively much larger than that of HGY. Since PBE exchange-correlation underestimates the band gap, we have also calculated the band gap of HGY and BN-HGY structures by employing HSE06[81] functional. We found that the band gap of the HGY structure is 0.9 eV using HSE06 functional, excellently matching with the experimental value of 1 eV[79]. Here, we report 5.18 eV of band gap for BN-HGY structure using HSE06 exchange-correlation. BN analog of graphene and graphyne (BN-yne) are also direct band gap materials with band gaps 4.64 eV and 2.65 eV, respectively, computed using the PBE exchange-correlation[61,82]. Since nitrogen atom is more electronegative compared to boron, nitrogen and boron atoms are connected by strongly polarized covalent bond, due to which the BN structures are wide bandgap materials. The total density of states of the HGY and BN-HGY structure is symmetric for the upper and lower panels, as shown in **Fig. 5,** which explains the non-magnetic nature of these structures. Since the HGY and BN-HGY structures have sp and $sp^2$–hybridized carbon, boron and nitrogen atoms, we have plotted the partial density of states (PDOS) of the 2p-orbitals of carbon, boron, and nitrogen atoms as displayed in **Fig. 6 (a & b).**

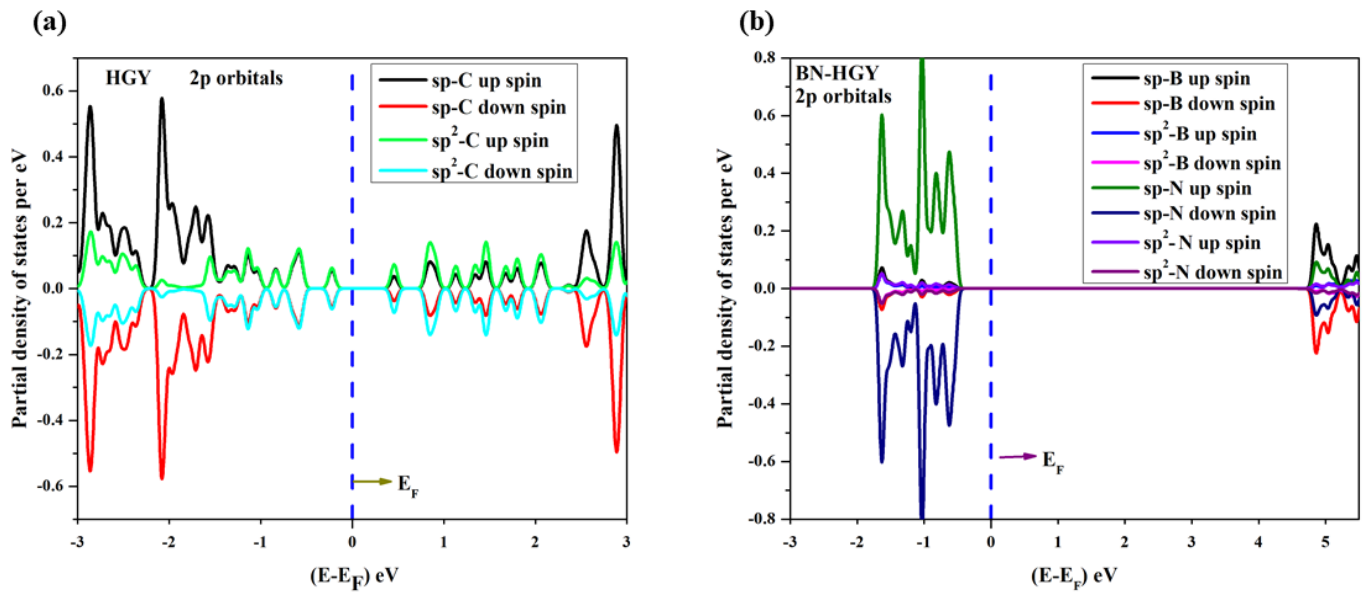

**Fig. 6 Partial density of states of 2p-orbitals of the (a) sp and $sp^2$-hybridized carbon atoms of holey graphyne (b) sp and $sp^2$-hybridized boron and nitrogen atoms of the boron nitride analog of monolayer holey graphyne. $E_F$ represents the Fermi level.**

The total density of states of the HGY structure is dominated by the PDOS of the 2p-orbitals of carbon atoms. We have observed more intense states for sp-hybridized carbon atoms compared to $sp^2$-hybridized carbon in HGY. The total density of states of BN-HGY structure is mainly contributed by the 2p-orbitals of the sp-hybridized nitrogen atoms. The sp-hybridized bond is more electronegative due to more contribution of s-orbitals as compared to $sp^2$-hybridization. This is the reason behind the intense peaks of the sp-hybridized carbon and nitrogen atoms in the 2p-orbitals PDOS of HGY and BN-HGY. We have plotted the detailed band structures of HGY and BN-HGY along the G-M-K-G high symmetry path, as shown in **Fig. 7,** and report that the BN-HGY structure is a direct band gap material.

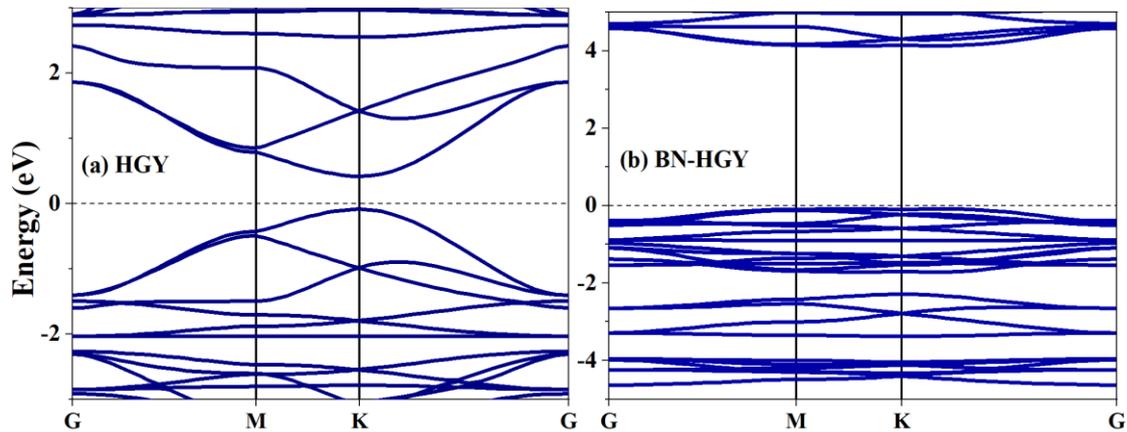

**Fig. 7 Band structure plots of (a) Monolayer holey graphyne (b) Boron nitride analog of holey graphyne.**

To further investigate the electronic properties of the BN-HGY structure, we have plotted the charge density plot for an isosurface value of 0.133e, as shown in **Fig. 8 (a).** The yellow color region around the nitrogen atoms denotes the charge accumulation region. We have also plotted the ELF for the accurate investigation of charge distribution in this material, as displayed in **Fig 8 (b).** The red color region of **Fig. 8 (b)** denotes the charge gain region where electrons are localized. We found that the electrons are localized around the nitrogen atoms of BN-HGY due to their high electronegativity compared to boron atoms. As electronic charge is accumulated around the nitrogen atoms, the covalent B-N bonds in this structure get polarized.

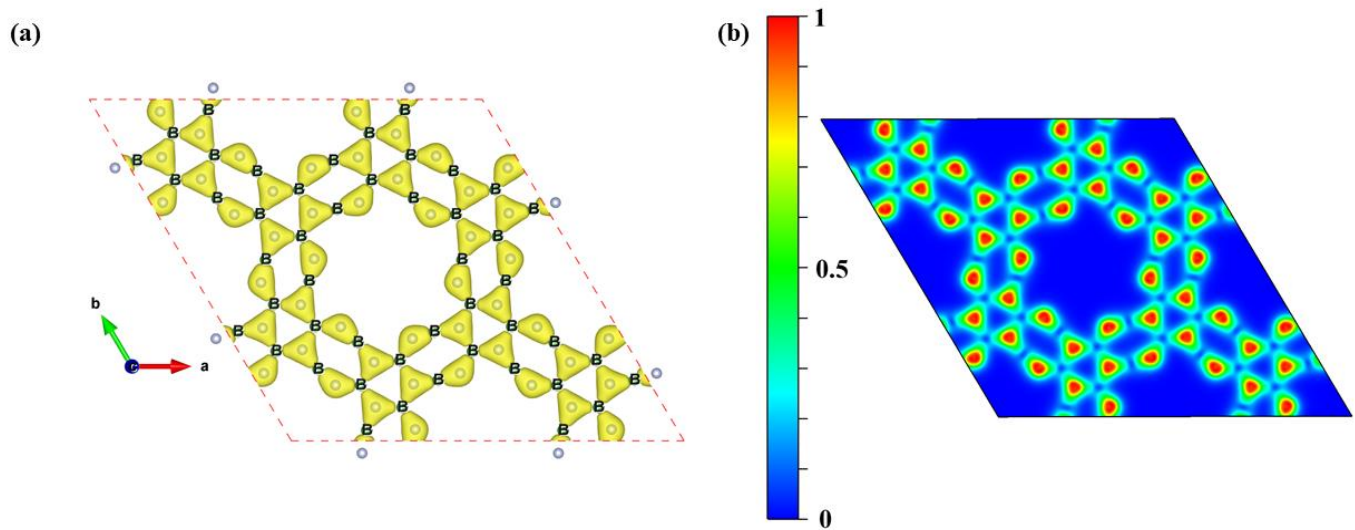

**Fig. 8 (a) Charge density plot of the BN-HGY structure. Here, yellow colored region around the nitrogen atoms denotes the charge accumulation region. (b) Electron localization function plot for the BN-HGY structure. Here, red colored region around the nitrogen atoms denotes the charge rich region.**

### 3.4 Bandgap engineering

Next, we have investigated the bandgap modulation in the BN analog of the HGY structure by substituting carbon (C), aluminum (Al), silicon (Si), and phosphorus (P) atoms in place of boron and nitrogen atoms and found that the electronic bandgap of the BN-HGY structure can be reduced significantly by doping C at the B1, N1 (sp-hybridized) and B2, N2 ($sp^2$ hybridized) boron and nitrogen atoms. The total density of states of the C doped BN-HGY structures is shown in **Fig. 9.** We found that the large band gap 4.22 eV of the BN-HGY structure can be tuned to 0.35 eV, by the doping of a single carbon atom at the sp-hybridized nitrogen atom, as shown in **Fig. 9 (c).**

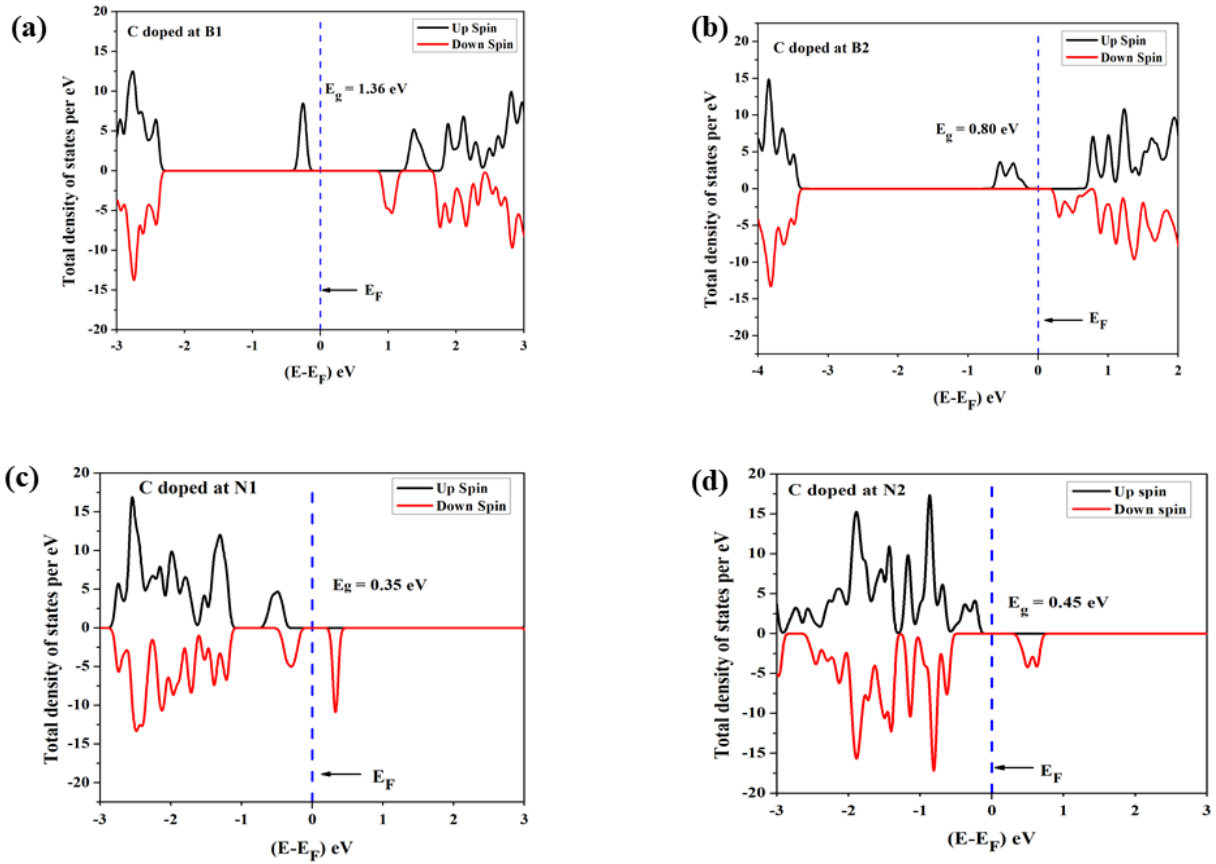

**Fig. 9** Total density of states of BN-HGY with (a) Single carbon atom doped at sp-hybridized boron atom (b) Single carbon atom doped at $sp^2$-hybridized boron atom (c) Single carbon atom doped at sp-hybridized nitrogen atom (d) Single carbon atom doped at $sp^2$-hybridized nitrogen atom. $E_F$ represents the Fermi level.

Further, we have substituted Al, Si, and P atoms at the sp and $sp^2$ hybridized boron and nitrogen atoms of the BN-HGY structure and found that the band gap of the pristine BN-HGY structure can be controllably tuned by the doping of single Al, Si, and P atoms at the sp and $sp^2$-hybridized boron atoms (B1, B2). The BN-HGY structure gets distorted when Al, Si, and P atoms are placed at N1 and N2. The total density of states of the Al, Si, and P doped BN-HGY structures are shown in **Fig. S3** of supporting information. Since the electronic band gap of the BN-HGY structure can be engineered in a controllable manner by the doping of C, Al, Si, and P atoms, the BN-HGY structure has potential applications in nanoelectronics. We have also observed that the electronic bandgap of the BN-HGY structure reduces significantly by light alkali, alkaline-earth, and transition-metal (Li, Mg, Sc, and Y) decoration, which indicates that the porous BN-HGY structure can be utilized for energy storage purposes. Further, we have found that the non-magnetic pristine BN-HGY structure becomes magnetic with the doping of C, Si, and P atoms and the decoration of Li, Sc, and Y metal atoms suggesting that the substituted BN-HGY structure can be utilized for spintronic applications. The bandgap values and induced magnetic moment of the C, Al, Si, and P doped and some metal-decorated BN-HGY structures are presented in **Table 2.**

**Table 2. Tunable electronic band gaps and induced magnetic moment of the BN-HGY structure by doping of C, Si, Al, and P atoms, and decorating some light metal (Li, Mg, Sc, Y) atoms.**

| Structure | Electronic bandgap (eV) PBE exchange-correlation | Magnetic moment (µB) |
|---|---|---|
| **Pristine BN-HGY** | 4.22 | 0.00 |
| **C doped at B1** | 1.36 | 1.00 |
| **C doped at B2** | 0.80 | 1.00 |
| **C doped at N1** | 0.35 | 1.00 |
| **C doped at N2** | 0.45 | 1.00 |
| **Al doped at B1** | 2.11 | 0.00 |
| **Al doped at B2** | 4.00 | 0.00 |
| **Si doped at B1** | 0.97 | 1.00 |
| **Si doped at B2** | 1.37 | 1.00 |
| **P doped at B1** | 2.37 | 0.00 |
| **P doped at B2** | 0.82 | 2.00 |
| **Li decorated BN-HGY** | 0.97 | 1.00 |
| **Mg decorated BN-HGY** | 2.44 | 0.00 |
| **Sc decorated BN-HGY** | 0.51 | 1.00 |
| **Y decorated BN-HGY** | 0.48 | 1.00 |

The total density of states for Li, Mg, Sc, and Y decorated BN-HGY structure is presented in **Fig. S4** of the supporting information.

### 3.5 Optical properties

In order to investigate the suitability of the HGY and BN-HGY structures for optoelectronics, we have calculated the dielectric functions, refractive indices, and absorption coefficients of these materials. To the best of our knowledge, no such previous calculations exist even for HGY. We have plotted the real and imaginary parts of the dielectric functions of monolayer HGY and BN-HGY structures with respect to energy, as shown in **Fig. 10**. The linear optical absorption can be calculated by computing the frequency-dependent complex dielectric function $\varepsilon(\omega) = \varepsilon_1(\omega) + i\,\varepsilon_2(\omega)$, where $\varepsilon_1(\omega)$ and $\varepsilon_2(\omega)$ are its real and imaginary parts, respectively.

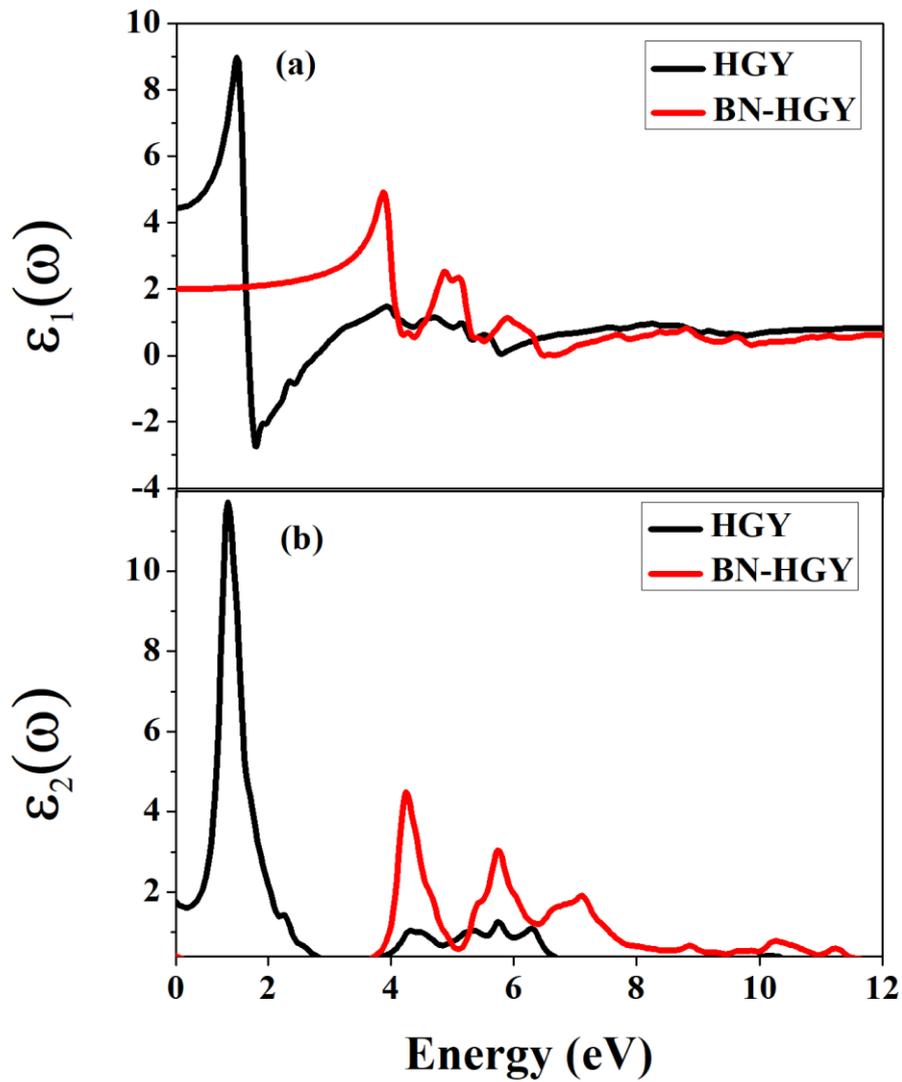

**Fig. 10 Variation of the real and imaginary parts of the dielectric function of HGY and BN-HGY structure with respect to the energy of the incident photon.**

The information regarding the response of a material to an incident external electromagnetic radiation of frequency ω is contained in $\varepsilon(\omega)$. We have also calculated the refractive index ***n(ω)*** and absorption coefficient ***α(ω)*** of these materials by using the following equations:

$$n(\omega) = \frac{1}{\sqrt{2}} \left[ \sqrt{\varepsilon_1^2 + \varepsilon_2^2} + \varepsilon_1 \right]^{\frac{1}{2}} \quad (1)$$

$$\alpha(\omega) = \sqrt{2}\,\omega \left[ \sqrt{\varepsilon_1^2 + \varepsilon_2^2} - \varepsilon_1 \right]^{\frac{1}{2}} \quad (2)$$

The frequency dependent linear optical absorption coefficients of the monolayer HGY and BN-HGY structures are presented in **Fig. 11,** while the refractive index ***n (ω)*** of the BN-HGY structure is presented in **Fig. S5** of the supporting information.

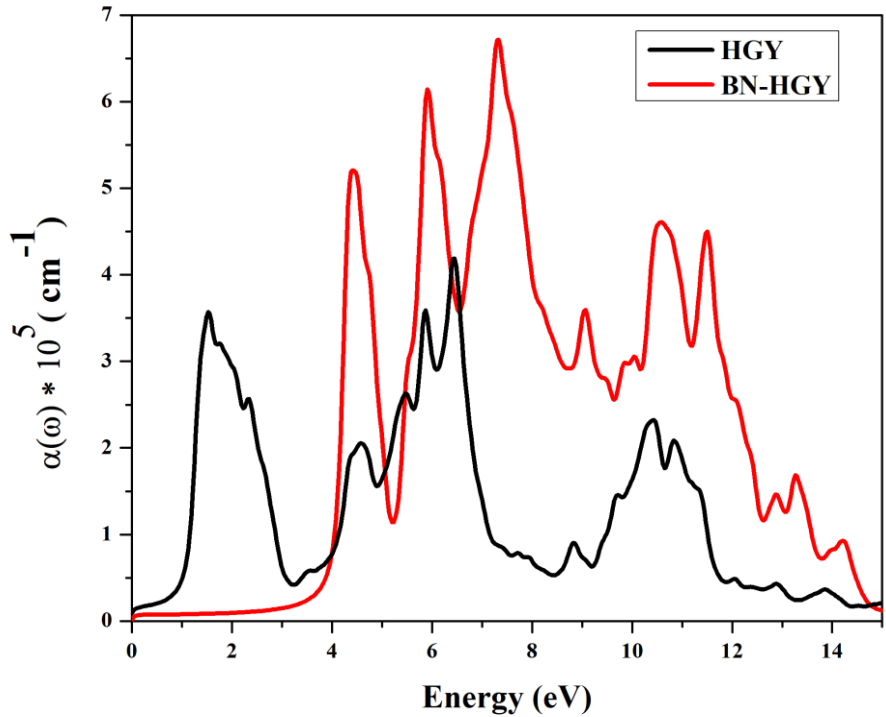

**Fig. 11 Optical absorption coefficient of the HGY and BN-HGY structures, plotted with respect to the energy of the incident photon.**

We have found that the linear optical absorption spectrum of HGY covers a wide range of visible and ultraviolet regions, while the optical absorption spectrum of the BN-HGY structure spans a wide range of ultraviolet region. Since the optical absorption of the HGY covers full visible region and a wide range of ultraviolet region, it will have potential applications in optoelectronics and photovoltaic solar cells. The optical absorption of the BN-HGY structure span a wide range of ultraviolet regions and it will have potential applications in ultraviolet laser devices and it can be utilized for photocatalysis and optical storage applications.

### 3.6 Catalytic performance for HER activity

We have also investigated the catalytic performance of the BN-HGY structure towards the hydrogen evolution reaction (HER). The HER performance of a material can be estimated by calculating its Gibbs free energy ($\Delta G$), which is given by the following equation:

$$\Delta G = \Delta E_{ads} + \Delta E^{ZPE} - T\Delta S \qquad (3)$$

Here $\Delta E_{ads}$ is the adsorption energy of the hydrogen atom on the surface of BN-HGY, $\Delta E^{ZPE}$ is the zero-point energy difference for the adsorbed hydrogen atom and isolated hydrogen in the gas phase, and $T\Delta S$ is the entropy correction term. The values of zero-point energy difference and entropy correction terms are taken from the literature[83,84]. Catalyst with $\Delta G$ close to zero is considered suitable for HER activity due to the feasibility of the hydrogen adsorption and desorption process. We have found that the adsorption energy of hydrogen is -0.03 eV and -0.28 eV when it is kept on the top of the center of the hexagon (H) and octagon (O) sites of the BN-HGY structure, respectively. The bond length of the adsorbed hydrogen and the nearest atom of BN-HGY is 3.82 Å and 1.43 Å, when hydrogen is adsorbed at H and O sites, respectively. We present the reaction coordinates plot for the HER activity of the BN-HGY structure in **Fig. 12.**

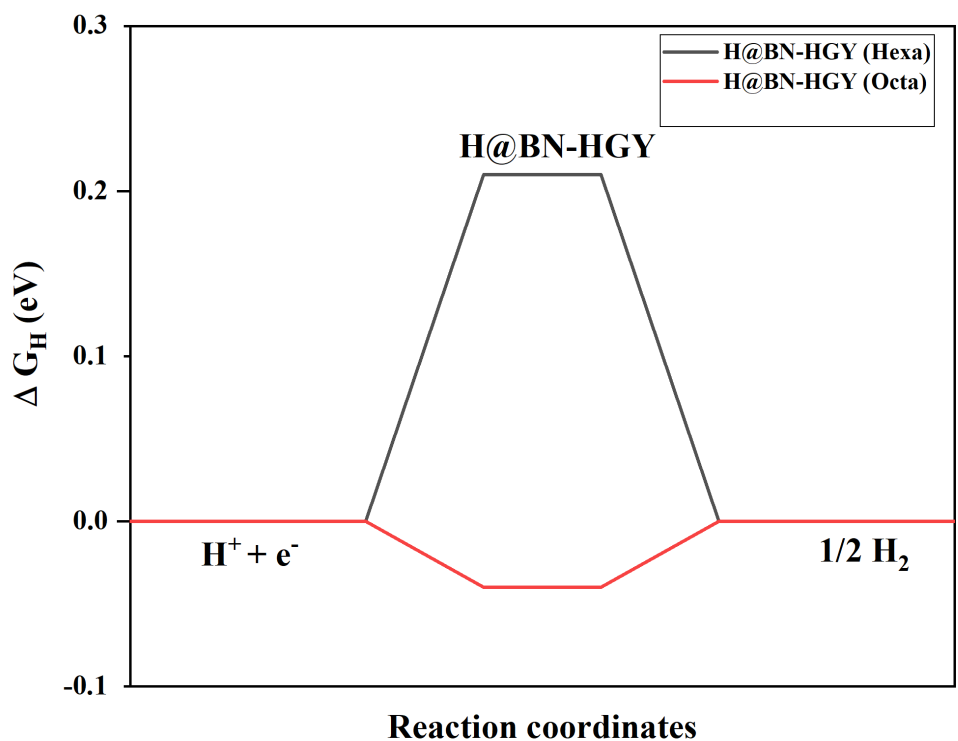

**Fig. 12 Reaction coordinate plot for the HER activity of pristine BN-HGY structure.**

We have found that the Gibbs free energy values of hydrogen adsorbed BN-HGY structure is 0.21 eV and -0.04 eV when hydrogen atom is attached to H and O sites, respectively which are very close to zero. Therefore, the pristine BN-HGY structure is an excellent catalyst for the HER activity.

4. **CONCLUSIONS**

In summary, we have theoretically predicted the novel BN-HGY structure and explored its various properties for the first time. We found that the BN-HGY structure is energetically stable, and can be realized experimentally. The dynamical and thermal stability of the BN-HGY structure is confirmed by calculating the phonon spectra and performing the *ab-initio* molecular dynamics simulations, respectively. We have observed that the BN-HGY structure is stable at room temperature and can withstand up to 1000 K. The boron nitride analog of HGY is a wide bandgap structure with a direct gap of 5.18 eV calculated using HSE06 exchange-correlation. We have found that the bandgap of the BN-HGY structure can be tuned in a controllable manner and the non-magnetic BN-HGY structure becomes magnetic with the doping of carbon, aluminum, silicon, and phosphorus atoms in place of boron and nitrogen atoms and decoration of light metal atoms, therefore, it will have potential applications in electronic and spintronic devices. The linear optical absorption spectra of pristine HGY and BN-HGY structures span a wide range of visible and ultraviolet regions, respectively, making them applicable for solar cell, ultraviolet laser, and optoelectronic device fabrication. The Gibbs free energy value for the hydrogen adsorption to the octagon site of BN-HGY is -0.04, which is very suitable for a potential catalyst for HER activity. We propose that the BN-HGY structure can be synthesized and utilized in nanoelectronics, spintronics, optoelectronics, and, catalyst for hydrogen evolution reaction.

## Conflicts of Interest

There are no conflicts of interest to declare.

**ACKNOWLEDGEMENTS**

V. M. acknowledge Department of Science and Technology (DST), New-Delhi, for providing DST-INSPIRE fellowship. V.M. would also like to acknowledge SpaceTime-2 supercomputing facility at IIT Bombay for the computing time. BC acknowledges Dr. Nandini Garg, Dr. T. Sakuntala, Dr. S.M. Yusuf and Dr. A.K. Mohanty for support and encouragement.

(X=N, P, As, Sb, Bi) Single-Atom catalyst: Ultra-high performance for hydrogen evolution reaction, *Int. J. Hydrogen Energy*, , DOI:10.1016/j.ijhydene.2022.02.159.

# Supporting Information

# Prediction of a Novel 2D Porous Boron Nitride Material with Excellent Electronic, Optical and Catalytic Properties

*Vikram Mahamiya[a], Alok Shukla[a*], and Brahmananda Chakraborty[b,c]*

[a]Indian Institute of Technology Bombay, Mumbai 400076, India

[b]High pressure and Synchrotron Radiation Physics Division, Bhabha Atomic Research Centre, Bombay, Mumbai, India-40085

[c]Homi Bhabha National Institute, Mumbai, India-400094

email: mahamiyavikram@gmail.com ; shukla@iitb.ac.in ; brahma@barc.gov.in

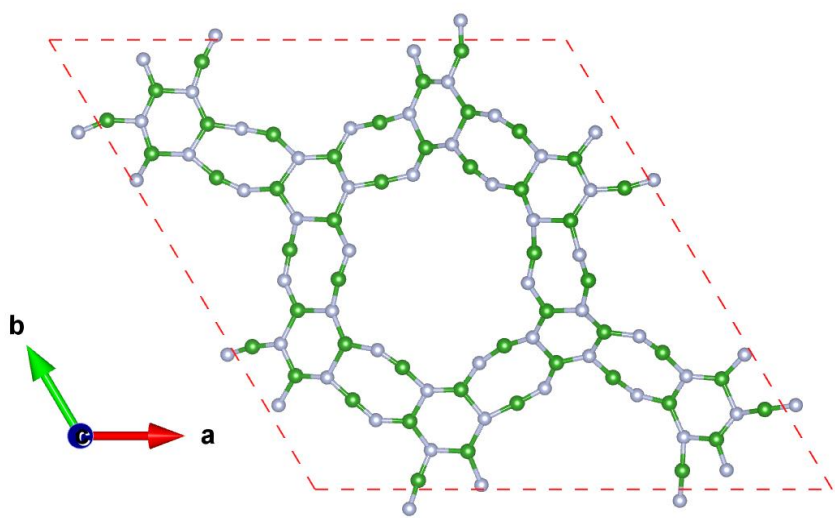

**Fig. S1** Molecular dynamics snapshot of the monolayer BN-HGY after keeping the structure in a canonical ensemble for 5 ps time duration at 1000 K.

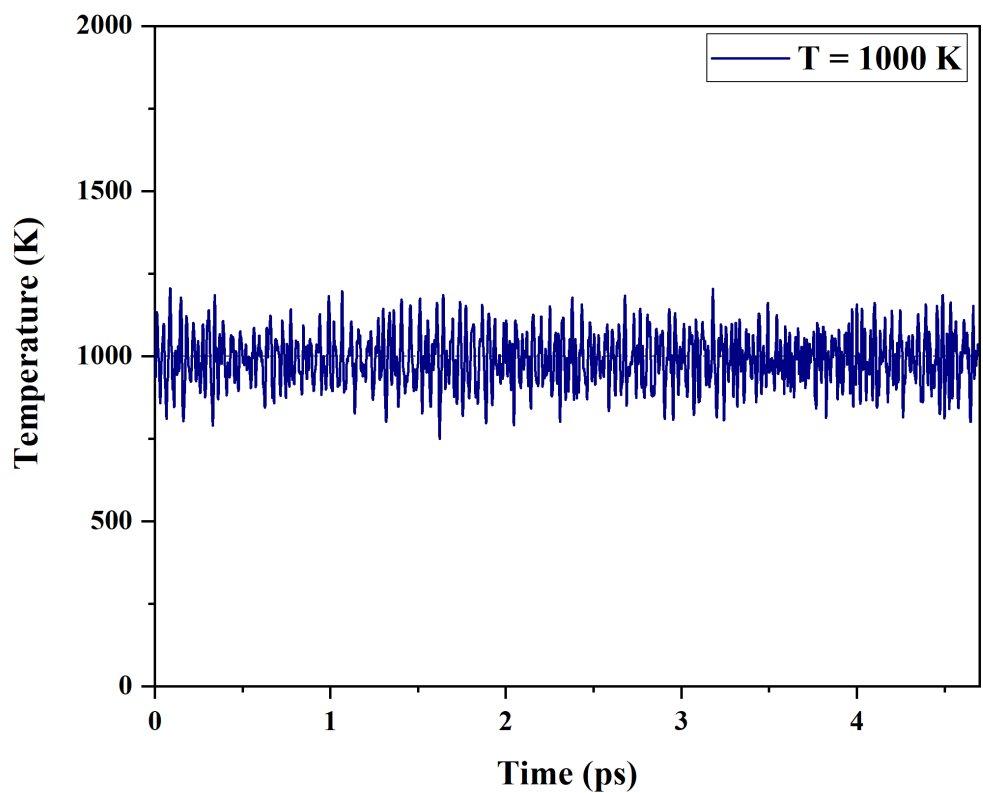

**Fig. S2** Temperature fluctuations in the monolayer BN-HGY after keeping the structure in a canonical ensemble for 5 ps time duration at 1000 K.

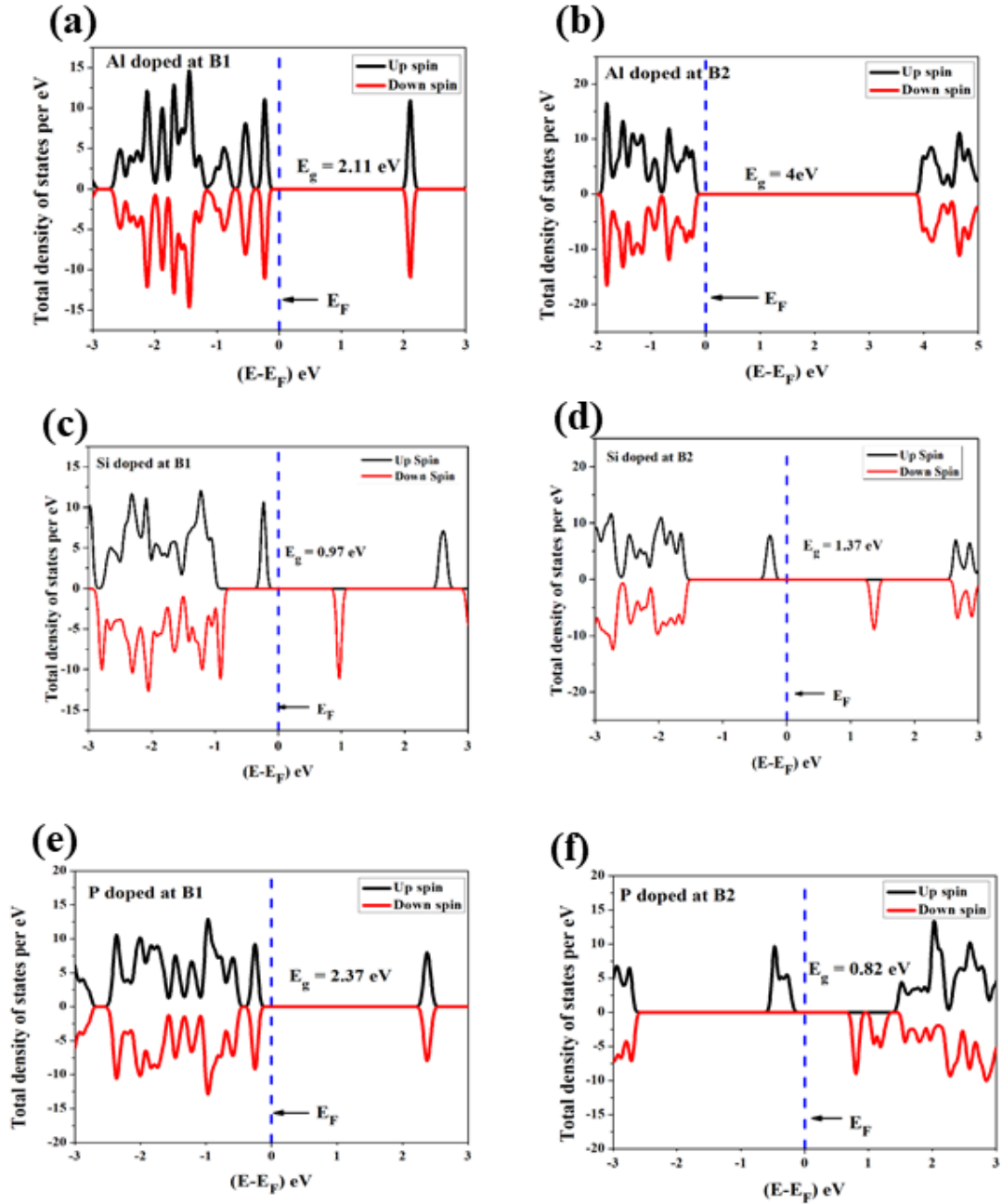

Fig. S3 Total density of states of BN-HGY with (a) Single aluminum atom doped at sp-hybridized boron atom (b) Single aluminum atom doped at $sp^2$-hybridized boron atom (c) Single silicon atom doped at sp-hybridized boron atom (d) Single silicon atom doped at $sp^2$-hybridized boron atom (e) Single phosphorus atom doped at sp-hybridized boron atom (f) Single phosphorus atom doped at $sp^2$-hybridized boron atom. $E_F$ represents the Fermi level.

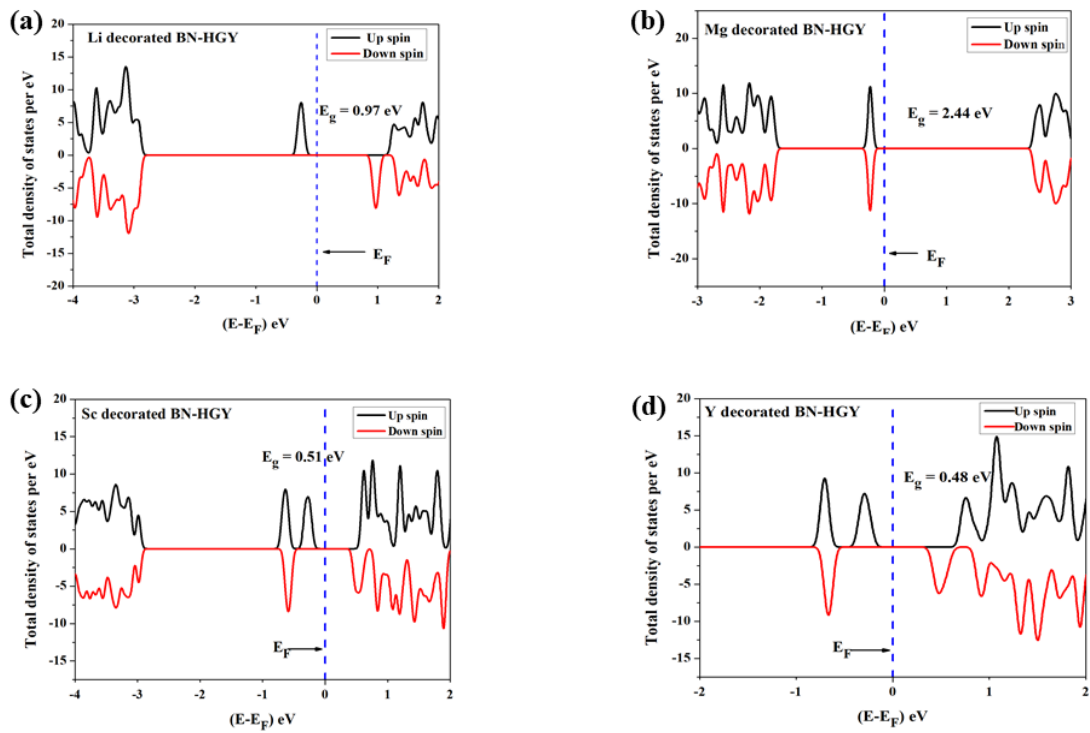

**Fig. S4** Total density of states of (a) Li decorated BN-HGY (b) Mg-decorated BN-HGY (c) Sc-decorated BN-HGY (d) Y-decorated BN-HGY. $E_F$ denotes Fermi energy.

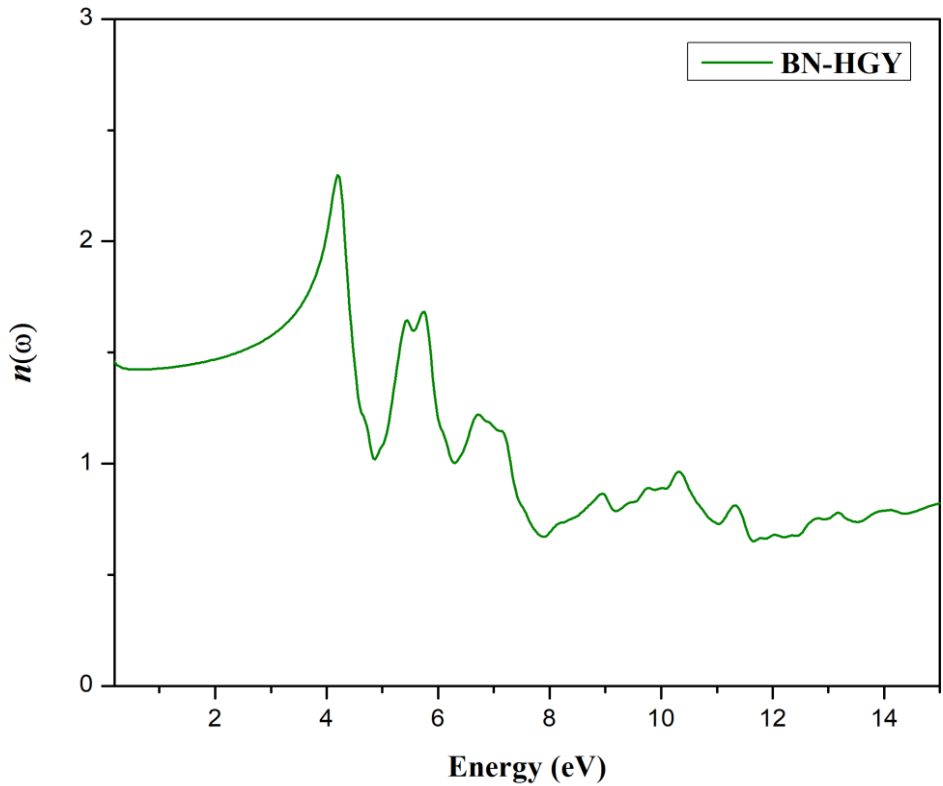

**Fig. S5** The refractive index plot of the BN-HGY structure with respect to the energy of incident photon.

**Structural files:**

1. Position of atoms in BN-HGY unit cell

BN-HGY

  1.00000000000000

    10.9187088012999993   0.0000000000000000   0.0000000000000000

   -5.4644601315000001   9.4615217934999993   0.0000000000000000

    1.3044178647000000   0.7099982333000000   7.5775218135999998

  B   N

  12   12

Direct

| | | |
|---|---|---|
| 0.0566971741682597 | 0.6389019931994668 | 0.5018388516514525 |
| 0.1926874521823643 | 0.5241284730690586 | 0.5012197681296378 |
| 0.3218171424157124 | 0.7982716503003187 | 0.5016855801975619 |
| 0.4669220795893775 | 0.6691744218433832 | 0.5018387624006801 |
| 0.5725211902550887 | 0.9341637378051905 | 0.5019189730254173 |
| 0.3519706552889871 | 0.4184862157155985 | 0.5012359759626668 |
| 0.6353967778918047 | 0.5757821353811766 | 0.5022334759707305 |
| 0.5204436674834406 | 0.3250939335024693 | 0.5015809636535463 |
| 0.7947645531143873 | 0.4701586566270122 | 0.5016157812398711 |
| 0.6655779729666641 | 0.1960166681748829 | 0.5016103778809397 |
| 0.4147832580016411 | 0.0600643152078548 | 0.5018746838700133 |
| 0.9307550666637444 | 0.3553966042863624 | 0.5010272359740948 |
| 0.3226075156284158 | 0.9275871673930171 | 0.5018549750617227 |
| 0.1939316097735876 | 0.6610441430644834 | 0.5014884701381875 |
| 0.0632850876362697 | 0.3956218344595660 | 0.5009678428583562 |
| 0.4575531243243695 | 0.7970143007302519 | 0.5017688523523787 |
| 0.3300044546311929 | 0.5335609467509147 | 0.5011614047239463 |
| 0.3919623584420341 | 0.3260800870234892 | 0.5013022215545587 |
| 0.5953149394131602 | 0.6681437310394783 | 0.5027267575151194 |
| 0.6574528424741708 | 0.4607432088436683 | 0.5017602759131905 |
| 0.5297974545867215 | 0.1972455870458008 | 0.5016512609867482 |
| 0.7935621599514922 | 0.3332835057476856 | 0.5011492330039808 |
| 0.6646755723940350 | 0.0666394356041891 | 0.5020633883436865 |
| 0.9241158907230623 | 0.5986472471846720 | 0.5021248875915024 |

## 2. BN-HGY (2*2 supercell) coordinates after keeping the system in canonical ensemble for 5 ps at 1000 K

BN-HGY

 1.00000000000000

  21.8331655023710773  -0.0047829008094971   0.4308794721713001

 -10.9265132170076207  18.9244222155918465  -0.0197986828702134

   1.1546515469311316   0.6312696178755490   7.6087265711826575

  B  N

  48  48

Direct

 0.2036281492551567  0.0280764654485088  0.5710222100559605

 0.2678061777307291  0.4643846405387267  0.6190915756682150

 0.3050929325133986  0.2802539647389553  0.6109439486508895

 0.1593161314111343  0.2034104530422128  0.6294810597982207

 0.4601938521841759  0.1699757458804616  0.5871644829483099

 0.0120527634950182  0.3207585580176112  0.5451805408171497

 0.0779108853492657  0.2655464486561769  0.6175311741872769

 0.3998220157295607  0.2337576299447987  0.6082082295976003

 0.2171669001141310  0.3288617985019516  0.6211106889710546

 0.2464222545365977  0.1593845209284770  0.6088194660204741

 0.1439034117663819  0.3988898743237075  0.6343725685998401

 0.3267680840811488  0.0899948261509044  0.5984998634633708

 0.7166385103849671  0.0303083269181663  0.3885215117849110

 0.7910841738411020  0.4566833139316616  0.4373179960059701

 0.8105390596525683  0.2826980347021949  0.4124203619663437

 0.6773419015502127  0.2076077775290202  0.4228284503828545

0.9563639452784801	0.1822345783922670	0.5333618461010741
0.5369227819712191	0.3259767688038166	0.5080955701484026
0.5935168386436719	0.2610353533244100	0.4899391430391826
0.8836183682436183	0.2324242740242426	0.5019601770464384
0.7334086942150361	0.3329624053219402	0.4163936018948027
0.7540137216275319	0.1588636390854377	0.4415285542885732
0.6700803072694629	0.4040265673451561	0.4266473233462995
0.8283295340615343	0.0973745342867041	0.4821222599525817
0.1993160619447869	0.5366471748893612	0.5965399197919804
0.2818007919881005	0.9690326093352192	0.5236486860985999
0.3379264833966513	0.7980946130787244	0.4718321772605788
0.1806219789210946	0.7133700955443698	0.4889208536587170
0.4678930938348433	0.6843672120911711	0.4860767644415845
0.0261931503448799	0.8196054729276446	0.5132817118855684
0.1031525182473200	0.7727910104485824	0.5238412446345054
0.4026791851676162	0.7339669915735516	0.5004788752570800
0.2425357693111955	0.8405513129005128	0.4698954114118827
0.2702225984302101	0.6668874055367587	0.5195672250911293
0.1617921287580946	0.9019915949527576	0.5223868605061228
0.3318491250544859	0.5947782095120985	0.5369980703645492
0.7211010632045479	0.5309377903719137	0.3868169994027751
0.7959184592642345	0.9677532589246399	0.4011122644955291
0.8153038978060662	0.7796054386777722	0.4999613794502541
0.7023844501296874	0.7223119957831291	0.4007379470945192
0.9698856280704190	0.6795294244330000	0.5233612824391263
0.5385562277904894	0.8195846490456975	0.4070311773929284
0.6056450004361985	0.7653443190382317	0.4116162431847509

0.9017304695747720  0.7304829373608045  0.5062550811679223
0.7417404683796663  0.8383660806204103  0.4077831624862359
0.7658991319478162  0.6634139814078596  0.4538821936948894
0.6802002112201766  0.9087430610377327  0.3506472512050274
0.8383697091033305  0.5947159541810579  0.4684711236184836
0.0153956097870634  0.1974676057651645  0.5947576624195341
0.4686583988698264  0.2971631219919850  0.5517869623349980
0.2820900497344097  0.3257419007238896  0.6103051647655964
0.1889093200137907  0.1650709332407357  0.6297515273231397
0.1474396258042908  0.4694400351645467  0.6328796622634102
0.3257656988528185  0.0285587711800658  0.5677925410850810
0.2631300448750821  0.0983484378263844  0.5846115743174616
0.2143805592776156  0.3932987063460976  0.6284915977622849
0.3195646972707713  0.2244333409886548  0.5984701707543658
0.1423189329494558  0.2604139244196826  0.6380410210483887
0.3910529921990516  0.1552382064059745  0.6214498414901723
0.0814217447813934  0.3339036297553277  0.5875900042216677
0.5272241856938954  0.2011426480365264  0.5331527857159237
0.9411043148225871  0.2961918429108191  0.5598022359974238
0.8004211439934038  0.3385184302155906  0.4116450248433894
0.6864404181515941  0.1498027394921835  0.4143850062697798
0.6714498967636030  0.4669204382384811  0.3716429886812222
0.8324527489103654  0.0316606901442585  0.4332302476433474
0.7602801096162659  0.0933050171181440  0.4502740378242267
0.7345144911888531  0.3957650748094211  0.4042882269277618
0.8200513380774878  0.2258565147317521  0.4519035572528164
0.6669411832989535  0.2654846919305127  0.4490628660357877

| | | |
|---|---|---|
| 0.8912937774837462 | 0.1661946483040107 | 0.4928303826769458 |
| 0.6059714785478597 | 0.3324662373282478 | 0.4555714297875061 |
| 0.0364542107150111 | 0.7091816637653520 | 0.5419229739128160 |
| 0.4716724720907199 | 0.7968039800097688 | 0.4873620884997042 |
| 0.3090749447115717 | 0.8388846087615381 | 0.4638091743327756 |
| 0.2163012328560122 | 0.6833068044031865 | 0.5031500889918384 |
| 0.1565279226595129 | 0.9639938516675695 | 0.5511525787796777 |
| 0.3276543182865906 | 0.5291712604957809 | 0.5580072114576948 |
| 0.2657780890316491 | 0.6008311579280678 | 0.5486610028869615 |
| 0.2312839662724321 | 0.9024749858968487 | 0.4798126677367953 |
| 0.3358159140402322 | 0.7364625145955531 | 0.5230261165617168 |
| 0.1722420740279486 | 0.7752418621797524 | 0.4718732742199275 |
| 0.3968312390887554 | 0.6651897036440303 | 0.5143875424483740 |
| 0.0976881362015598 | 0.8339424683567281 | 0.5248343746788446 |
| 0.5333539356731349 | 0.6994729611661658 | 0.4541491415263179 |
| 0.9592877294838226 | 0.7991912819069537 | 0.5037539231845954 |
| 0.7961457605525440 | 0.8261293558057264 | 0.4760753657337705 |
| 0.7154627528770752 | 0.6702389603460113 | 0.3804331732098948 |
| 0.6791645559650548 | 0.9676730653592902 | 0.3465987577570060 |
| 0.8261565676640602 | 0.5243414021486195 | 0.4535701213994284 |
| 0.7729699326212152 | 0.5946550314653117 | 0.4289543561955404 |
| 0.7491959258869550 | 0.9062434239060168 | 0.3607290026739423 |
| 0.8322263371987463 | 0.7233107100235863 | 0.5048168681136913 |
| 0.6806307651236162 | 0.7689013858097599 | 0.3617915266578634 |
| 0.9014567151302696 | 0.6607245912577789 | 0.5039292193314608 |
| 0.6099451668070720 | 0.8328739215196856 | 0.3602942163631202 |